# Enhancement and inhibition of light tunneling mediated by resonant mode conversion

Yaroslav V. Kartashov,[1,2,*] Victor A. Vysloukh,[1] and Lluis Torner[1]

[1]*ICFO-Institut de Ciencies Fotoniques, and Universitat Politecnica de Catalunya, 08860 Castelldefels (Barcelona), Spain*
[2]*Institute of Spectroscopy, Russian Academy of Sciences, Troitsk, Moscow Region, 142190, Russia*

We show that the rate at which light tunnels between neighboring multimode waveguides can be drastically increased or reduced by the presence of small longitudinal periodic modulations of the waveguide properties that stimulate resonant conversion between eigenmodes of each waveguide. Such a conversion, available only in multimode guiding structures, leads to periodic power transfer into higher-order modes, whose tails may considerably overlap with neighboring waveguides. As a result, the effective coupling constant for neighboring waveguides may change *by several orders of magnitude* upon small variations in the longitudinal modulation parameters.

Tailoring the coupling of guided light modes is central to integrated-optics and fiber-optics technologies and devices, thus it is a topic of continuously renewed interest. Recent attention has focused on modulated multimode waveguides and, more generally, in microstructured materials and photonic crystals. In particular, it was shown that a longitudinal modulation of the waveguide depth or a harmonic bending of its axis may lead to resonant, Rabi-like oscillations and adiabatic transitions between guided modes (see [1], for a recent review, and references therein). Such transitions cause periodic oscillations of the weights of the modes when the propagation constant difference matches the frequency of the refractive index modulation.

Importantly, efficient mode conversion can occur not only in simple isolated waveguides, but also in complex guiding structures [2,3]. In particular, a rich set of phenomena can be generated when several longitudinally modulated waveguides are placed close to each other, so that light may tunnel between them. The tunneling rate is strongly affected by any periodic variations of the waveguide parameters. Thus, periodic bending of waveguides yields linear [4-9] or nonlinear [10,11] dynamic localization if the frequency and amplitude of bending are properly chosen, so that the band of propagation constants of collective modes shrinks. Periodic out-of-phase modulation of the depth or width of neighboring waveguides may lead to inhibition of tunneling for properly selected modulation frequencies, at which the effective coupling constant nearly vanishes [12-16].

To date, studies of such collective resonant effects have been focused on waveguiding structures consisting of the simplest elements, namely single-mode waveguides. However, in multimode waveguides the rate at which the excitations can expand across the structure may strongly depend on the particular modes that are excited [17]. Since resonant mode coupling leads to notable dynamical modifications of the mode weights, this effect may be used for control (enhancement or inhibition) of the global diffraction pattern of light in complex arrays of multimode waveguides.

In this Letter we consider one- and two-dimensional guiding structures composed of evanescently coupled multimode waveguides, whose parameters vary in a periodic fashion along the direction of light propagation. We study how a nearly resonant modulation of the waveguide depths stimulates coupling of the modes, thus drastically altering the tunneling rate between neighboring waveguides. Changes in coupling length by more than three orders of magnitude upon slight modifications of the longitudinal modulation frequency are shown. In contrast to guiding structures composed of single-mode waveguides, in the system under study, not only inhibition, but also enhancement of tunneling becomes possible.

We consider the propagation of paraxial light beams along the $\xi$ axis of a transparent optical medium with transversally and longitudinally inhomogeneous refractive index. The evolution of the dimensionless amplitude of the light field $q$ can be described by the equation:

$$i\frac{\partial q}{\partial \xi} = -\frac{1}{2}\Delta q - pR(\mathbf{r},\xi)q \qquad (1)$$

Here $\xi$ is the normalized propagation distance; $\mathbf{r}=\{\eta,\zeta\}$ is the normalized transverse coordinate; the Laplacian has the form $\Delta = \partial^2/\partial\eta^2 + \partial^2/\partial\zeta^2$ in the two-dimensional (2D) case and $\Delta = \partial^2/\partial\eta^2$ in the one-dimensional (1D) case; $p$ is the waveguiding parameter; and the function $R(\mathbf{r},\xi)$ describes the refractive index profile that varies periodically in the longitudinal direction $\xi$. Here we consider two types of the longitudinal refractive index variation: modulation of the waveguide depths, and periodic bending of their axes, which in the 1D case are described by the expressions:

$$R(\eta,\xi) = \sum_{j=1}^{n}[1+(\pm 1)^j \mu\sin(\Omega\xi)]\exp[-(\eta-jd)^4/a^4],$$
$$R(\eta,\xi) = \sum_{j=1}^{n}\exp\{-[\eta-(\pm 1)^j\mu\sin(\Omega\xi)-jd]^4/a^4\}. \qquad (2)$$

We assume that the refractive index landscape may include up to $n$ super-Gaussian waveguides of width $a$, separated by a distance $d$. The parameter $\mu$ determines the depth of the modulation (i.e., either the longitudinal refractive index modulation or the amplitude of waveguide bending), and $\Omega$ is the spatial modulation frequency. The parameters of the waveguides can oscillate in-phase [upper signs in Eq. (2)] or out-of-phase [lower signs in Eq. (2)]. The expressions for a 2D refractive index landscape $R(\eta,\zeta,\xi)$, composed of waveguides with the shape $\exp[-(\eta^2+\zeta^2)/a^2]$, are qualitatively similar. For $r_0=10\,\mu\text{m}$ and light beams at the wavelength $\lambda=633$ nm propagating in fused silica [6], a waveguide depth $p=10$ corresponds to a refractive index contrast $\delta n \sim 7\times 10^{-4}$, and $\xi=1$ corresponds to a propagation distance $\sim 1.4$ mm.

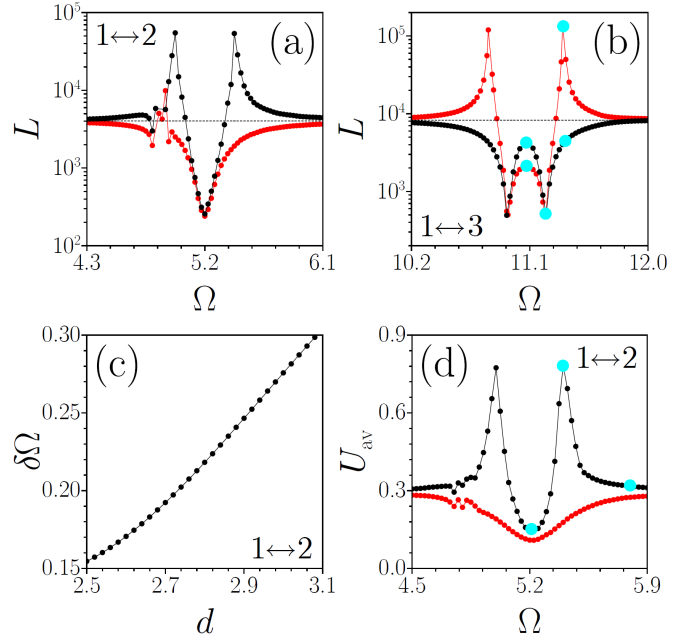

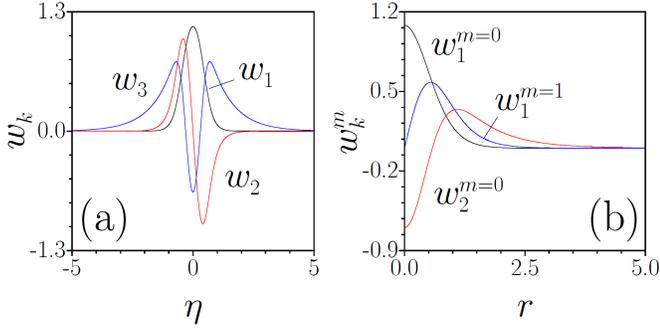

Fig. 1. Eigenmodes supported by (a) a 1D waveguide with $p=14$, $a=0.7$ and (b) a 2D waveguide with $p=10$, $a=1$. In (b) the superscripts $m$ stands for the winding number of the mode, while the subscripts $k$ carries the information on the number of nodes given by $k-1$.

We set the parameters $p,a$ in such a way that each waveguide supports several modes $[q=w_k(\eta)\exp(ib_k\xi)$ in the 1D case and $q=w_k^m(r)\exp(im\phi+ib_k\xi)$ in the 2D case] with different propagation constants $b_k$, different number of nodes $k-1$, and different winding numbers $m$ (available only for 2D modes). Figure 1 shows possible eigenmodes of static ($\mu=0$) isolated 1D and 2D waveguides that will be used as the input conditions in Eq. (1). The width of the mode notably grows with increase of its index $k$ (or vorticity $m$), with the fundamental mode being the most localized. In the presence of neighboring waveguides, the width of the modes is a key factor that determines the rate at which energy flow $U=\int|q|^2\,d\eta d\zeta$ tunnels from the excited waveguide into the neighboring ones. Thus, in the 1D case, in a static two-waveguide configuration with separation $d=3$ between channels, the power completely tunnels from the excited waveguide into the second one at a distance $L\approx 16500$ if the fundamental mode ($k=1$) is used as excitation at $\xi=0$, and at a distance $L\approx 21.9$ if the third mode ($k=3$) is used instead. Such a drastic contrast in coupling lengths suggests that controllable transfer of power between $k=1$ and $k=3$ modes should result in drastic modifications of the tunneling rate in a multi-waveguide system.

Fig. 2. Coupling length in a two-channel 1D structure versus $\Omega$ (a) for periodically bending waveguides with $p=12$, $d=3$ and (b) for waveguides with oscillating depths with $p=14$, $d=3$. (c) $\delta\Omega=\Omega_{\text{it}}-\Omega_r^{1\leftrightarrow 2}$ versus spacing between bending waveguides $d$ at $p=12$. (d) $U_{\text{av}}$ versus $\Omega$ for an array of bending waveguides with $p=12$, $d=2.6$. In (a),(b), and (d) black/red curves correspond to in-phase/out-of-phase modulation of the depth or position of neighboring channels. Dashed lines in (a),(b) show coupling length for fundamental mode in the unmodulated system. Blue circles in (b) and (d) correspond to dynamics shown in Figs. 3(a)-3(f) and 3(g)-3(i), respectively. In all cases $\mu=0.01$.

Resonant mode conversion allows achieving power transfer between modes of various parity. In the particular case of single 1D waveguide with oscillating depth [which corresponds to the first expression in Eq. (2)] one can show using standard coupled-mode approach [2] that if the combination of modes $q=\sum C_k w_k$ with weights $C_k(\xi)$ is used to excite the waveguide at $\xi=0$, then the mode weight evolution is governed by the equations $2dC_n/d\xi=\mu p\mathcal{I}C_m e^{i(b_m-b_n+\Omega)\xi}$, $2dC_m/d\xi=-\mu p\mathcal{I}C_n e^{i(b_n-b_m-\Omega)\xi}$, where $\mathcal{I}=\int w_n R_{\xi=0} w_m d\eta$ is the exchange integral. In such a waveguide the conversion is possible between modes of equal parity for which $\mathcal{I}\neq 0$ [for example between first $w_1$ and third $w_3$ modes from Fig. 1(a)]. The energy exchange, that is periodic in $\xi$, is most efficient when the resonance condition $b_m-b_n=\Omega$ connecting propagation constants and the longitudinal modulation frequency is satisfied. The efficiency of conversion drops down when the frequency $\Omega$ is detuned from the resonant value $b_m-b_n$ (the width of resonance decreases when $\mu$ decreases). Similarly, periodic bending of the waveguides enables conversion between modes of different parities (such as the $1\leftrightarrow 2$ conversion between the first and second modes that occurs for $\Omega=b_1-b_2$). Since for finite $\mu$ all resonances have nonzero width, a small fraction of power is always transferred into leaky modes thus causing a slow power leakage from the guiding structure.

This process is strongly affected by the presence of a neighboring multimode waveguide that is also dynamically modulated along the $\xi$ axis. Due to the drastic difference of tunneling rates into neighboring waveguides for different modes, the fraction of power transferred into high-order modes due to resonant conversion starts switching rapidly between the waveguides, while tunneling for the power fraction contained in the fundamental mode occurs at much larger scales. This process is accompanied by the continuous modification (due to resonant conversion) of the fraction of modes that couple rapidly and slowly, which evolves at its own scale determined, among other factors, by the modulation parameter $\mu$. Leaky modes may also contribute to the exchange, but because in our setting they carry a very small fraction of the input power, this contribution is small. Since in the framework of the coupled-mode approach the dynamics of the system is described by four equations for the weights of the resonantly coupled modes that cannot be solved analytically, we resort to direct numerical solution of Eq. (1).

In all cases we excite only one of the waveguides at $\xi=0$ with its fundamental mode and vary the modulation frequency $\Omega$ around the resonant value corresponding to the $1 \leftrightarrow 2$ (in the case of bending) or $1 \leftrightarrow 3$ (in the case of depth modulation) transitions and given by the difference of propagation constants of eigenmodes: $\Omega_r^{1 \leftrightarrow 2} = b_1 - b_2$ and $\Omega_r^{1 \leftrightarrow 3} = b_1 - b_3$. Figures 2(a) and 2(b) show in logarithmic scale the obtained dependencies of the effective coupling length $L$ - determined as a distance at which nearly all power is transferred from the excited into the neighboring waveguide - on the longitudinal modulation frequency in two periodically curved and depth-modulated waveguides, respectively. In periodically curved waveguides the minimum of the coupling length is achieved exactly at the resonance frequency $\Omega_r^{1 \leftrightarrow 2} \approx 5.19$. The minimal coupling length $L_{\min}$ is much smaller than the coupling length for the fundamental mode in the unmodulated waveguides (shown by the horizontal dashed line). Thus, the longitudinal refractive index variation allows considerable tunneling enhancement.

There is a drastic difference between in-phase and out-of-phase bending regimes. While for the out-of-phase bending, the distance $L$ nearly monotonically decreases as $\Omega \rightarrow \Omega_r^{1 \leftrightarrow 2}$, for the in-phase bending the coupling distance rapidly grows (by several orders of magnitude in comparison with $L_{\min}$) at two frequencies $\Omega_{\rm it} = \Omega_r^{1 \leftrightarrow 2} \pm \delta\Omega$ located symmetrically around the resonance. This effect is somewhat similar to the inhibition of tunneling in a sense that waveguides become effectively isolated due to the weak ($\mu = 0.01$) longitudinal modulation. However, in our case the inhibition of tunneling occurs only for two frequencies, rather than for an infinite set of them as in single-mode waveguides (see [14] for details). Far from the resonance the effective coupling length for both types of bending asymptotically approaches the coupling length for the fundamental mode in the unmodulated system. A similar picture is encountered in waveguides with modulated depths around the resonant modulation frequency $\Omega_r^{1 \leftrightarrow 3}$ [in Fig. 2(b) $\Omega_r^{1 \leftrightarrow 3} \approx 11.04$]. However, now the most effective coupling occurs at two frequencies slightly shifted from the resonant one. They nearly coincide for the in-phase and out-of-phase modulations of the waveguide depths.

In complete contrast with the case of bending, inhibition of light tunneling in the depth-modulated waveguides requires out-of-phase oscillations of the refractive index in the neighboring channels. When the separation $d$ between two waveguides increases, the frequencies corresponding to the minimal coupling length gradually become closer to each other. In contrast, the frequencies $\Omega_{\rm it} = \Omega_r^{1 \leftrightarrow 3} \pm \delta\Omega$, at which tunneling inhibition occurs, shift farther from the resonant one when $d$ increases. A typical dependence $\delta\Omega(d)$ of the detuning required for the tunneling inhibition on separation $d$ is shown in Fig. 2(c) for the in-phase bending of two waveguides. The propagation dynamics in depth-modulated two-channel system is illustrated in Fig. 3, for several characteristic frequencies. Notice the enhancement of tunneling in panels (c),(d) and its inhibition (f) at properly selected frequency of the out-of-phase refractive index modulation [for the same frequency and in-phase modulation the coupling still takes place in (e)].

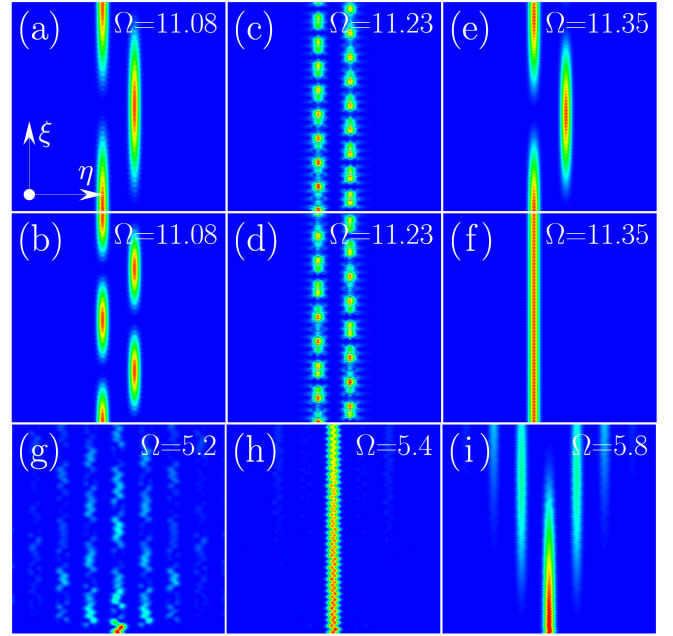

Fig. 3. Evolution dynamics in two-channel 1D system with $p=14$, $d=3$, and oscillating depths of the waveguides (a)-(f) and in the array with $p=12$, $d=2.6$, consisting of periodically bending waveguides (g)-(f) for different $\Omega$ values. Top row - in-phase refractive index modulation, middle row - out-of-phase modulation, while bottom row was obtained for the in-phase bending of channels. In all cases $\mu = 0.01$.

The above effects can be observed not only in two-waveguide structures, but also in arrays of multimode waveguides. Significantly different discrete diffraction rates for the fundamental and higher-order modes in the array make it difficult to determine the effective coupling length directly, therefore in this case it is more convenient to plot the dependence of the distance-averaged power $U_{\rm av} = L^{-1} \int_0^L d\xi \int_{-d/2}^{+d/2} |q|^2 \, d\eta$ in the excited channel on the modulation frequency [Fig. 2(d)]. This quantity characterizes the rate at which the excitation leaves the central channel, since a stronger diffraction corresponds to a smaller $U_{\rm av}$ value. The dependence $U_{\rm av}(\Omega)$ calculated for the array of curved waveguides around the $1 \leftrightarrow 2$ resonant transition contains all the characteristic features of the $L(\Omega)$ dependence depicted in Fig. 2(a), including resonant spikes for the in-phase bending. Figures 3(g)-3(i) illustrate enhanced tunneling, its inhibition, and off-resonant light propagation, respectively.

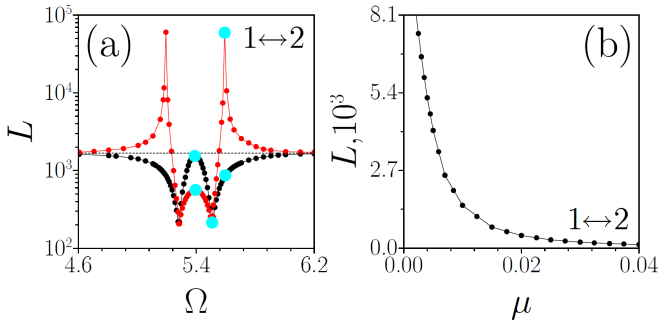

Fig. 4. (a) $L$ versus $\Omega$ in a two-channel 2D structure with $\mu=0.02$, $d=3.5$. Black/red curves correspond to the in-phase/out-of-phase modulation of the depth of neighboring channels. The dashed line shows the coupling length for the fundamental mode in unmodulated system. Blue circles correspond to evolution patterns from Fig. 5. (b) $L$ versus $\mu$ at $d=4$, $\Omega=5.406$ for in-phase refractive index modulation. In all cases $p=10$.

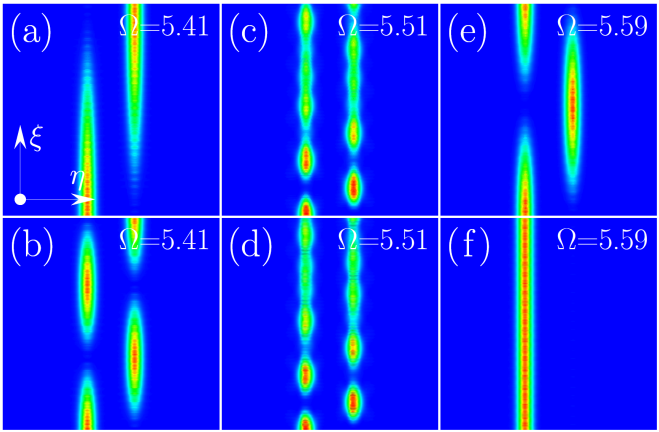

Fig. 5. Intensity distributions at $\zeta=0$ showing the light evolution in the $(\eta,\xi)$ plane in a two-channel 2D system with $p=10$, $d=3.5$, $\mu=0.02$, and oscillating depths of the waveguides for different modulation frequencies $\Omega$. Top row: in-phase refractive index modulation; bottom row: out-of-phase modulation.

The enhancement and inhibition of light tunneling mediated by resonant mode conversion occurs also in multidimensional geometries. The two-dimensional setting offers even more freedom, since resonant conversion may now occur between modes with different number of radial nodes [see Fig. 1(b)], and also between modes carrying different winding numbers $m$, as it was shown in [18-20]. In order to couple fundamental and vortex-carrying modes, the waveguide may be set into rotary motion with respect to an off-center axis. If such rotating waveguides are placed close to each other, one observes the whole set of resonance effects described above for the one-dimensional systems. Such effects, however, become more spectacular if resonant transition occurs between modes with different number of radial nodes, since for them the difference in widths is most pronounced. Figure 4(a) shows dependencies of coupling length on $\Omega$ in the system of two two-dimensional waveguides with in-phase and out-of-phase depth modulation in the vicinity of the frequency $\Omega_r^{1\leftrightarrow 2}\approx 5.41$ of resonant transition between $w_1^{m=0}$ and $w_2^{m=0}$ modes [Fig. 1(b)]. Like in the one-dimensional case, the enhancement of tunneling is possible for any type of modulation [see Figs. 5(c) and 5(d) corresponding to the minimal effective coupling length], while the inhibition of tunneling requires an out-of-phase refractive index oscillation in two waveguides [compare Figs. 5(e) and 5(f)]. Notice that the characteristic coupling length rapidly diminishes with increase of the depth of the longitudinal refractive index modulation [Fig. 4(b)].

Summarizing, we have shown how resonant mode conversion in longitudinally modulated multichannel guiding structures may efficiently enhance or strongly inhibit light tunneling between different waveguides. The coupling constant has been found to vary by orders of magnitude by small variations of the modulation parameters. The effects occur with modes of different internal structure, and can be sharply engineered. Similar effects may exist in nonlinear waveguides and lattices [15,21,22], where nonlinearity may cause notable asymmetries in resonant dependencies.